%% file: HIN-19-007_temp.tex
\pdfoutput=1
\documentclass[11pt,twoside,a4paper,cmspaper,final,collab]{cms-tdr}
\def\svnVersion{bb9fb52}\def\svnDate{2022/01/14}\def\cmsCernNoTag{CERN-EP-2021-104}\def\cmsCernDate{\today}\def\cmsMessage{Published in Physics Letters B as \href{http://dx.doi.org/10.1016/j.physletb.2021.136842}{\doi{10.1016/j.physletb.2021.136842}.}}
\begin{document}\cmsNoteHeader{HIN-19-007}

\newcommand{\sqrts}{\ensuremath{\sqrt{s}}\xspace}
\newcommand{\AuAu}{\ensuremath{\text{AuAu}}\xspace}
\newcommand{\PbPb}{\ensuremath{\text{PbPb}}\xspace}
\newcommand{\tAA}{\ensuremath{\mathrm{AA}}} 
\newcommand{\RAA}{\ensuremath{R_\tAA}\xspace}
\ifthenelse{\boolean{cms@external}}{\providecommand{\cmsLeft}{upper\xspace}}{\providecommand{\cmsLeft}{left\xspace}}
\ifthenelse{\boolean{cms@external}}{\providecommand{\cmsRight}{lower\xspace}}{\providecommand{\cmsRight}{right\xspace}}
\ifthenelse{\boolean{cms@external}}{\providecommand{\cmsLeftI}{Upper\xspace}}{\providecommand{\cmsLeftI}{Left\xspace}}
\ifthenelse{\boolean{cms@external}}{\providecommand{\cmsRightI}{Lower\xspace}}{\providecommand{\cmsRightI}{Right\xspace}}
\cmsNoteHeader{HIN-19-007}

\title{Fragmentation of jets containing a prompt \texorpdfstring{\JPsi}{Jpsi} meson in \texorpdfstring{\PbPb}{PbPb} and \texorpdfstring{$\Pp\Pp$}{pp} collisions at \texorpdfstring{$\sqrtsNN = 5.02\TeV$}{sqrt(s) = 5.02 TeV}}

\date{\today}

\abstract{
Jets containing a prompt \JPsi meson are studied in lead-lead collisions at a nucleon-nucleon center-of-mass energy of 5.02\TeV, using the CMS detector at the LHC.   Jets are selected to be in the transverse momentum range of $30 < \pt < 40\GeV$.   The \JPsi yield in these jets is evaluated as a function of the jet fragmentation variable $z$, the ratio of the \JPsi \pt to the jet \pt.  The nuclear modification factor, \RAA, is then derived by comparing the yield in lead-lead collisions to the corresponding expectation based on proton-proton data, at the same nucleon-nucleon center-of-mass energy.  The suppression of the \JPsi yield shows a dependence on $z$, indicating that the interaction of the \JPsi with the quark-gluon plasma formed in heavy ion collisions depends on the fragmentation that gives rise to the \JPsi meson.   
}

\hypersetup{%
pdfauthor={CMS Collaboration},%
pdftitle={Fragmentation of jets containing a J/Psi meson in lead-lead and proton-proton collisions at sqrt(sNN) = 5.02 TeV},%
pdfsubject={CMS},%
pdfkeywords={CMS, heavy ions, quarkonia, jets}}

\maketitle

\section{Introduction}
\label{sec:introduction}

Dissociation of quarkonium states in nucleus-nucleus collisions is one of the best studied signatures of the formation of the quark-gluon plasma (QGP), a deconfined state of quarks and gluons.  Although other nuclear effects have been identified, it is widely accepted that at least part of the suppression of the various quarkonium states in central lead-lead (\PbPb) collisions, at the collision energies probed at the CERN LHC, is indeed coming from Debye-like screening of heavy-quark pairs in the QGP, as anticipated in the seminal paper by Matsui and Satz~\cite{Matsui:1986dk}.  This picture may be probed with quarkonia produced at rest.  With the emergence of competing mechanisms, however, it becomes interesting to study the momentum dependence of the nuclear suppression.  In particular, this is true for regeneration~\cite{BraunMunzinger:2000px}, wherein quarkonia may be created from heavy quarks produced independently.  This effect is expected to become more relevant with increased collision energy, as more heavy-quark pairs are produced.   This explains the relatively low nuclear suppression at low transverse momentum (\pt) that is observed at the LHC~\cite{Abelev:2012rv}, as compared to data at lower collision energies~\cite{Abreu:2000ni,Adare:2011yf}.  Independent of these nuclear modifications, the interpretation of quarkonium results, and heavy-flavor results in general, is typically based on the assumption that quarkonia are formed at early times compared to the formation time of the QGP (on the order of 1\unit{fm}/$c$). 

However, the formation time estimate of quarkonia is based on general arguments rather than on a detailed calculation.  Despite decades of theoretical developments, models generally are not able to describe the entirety of the quarkonium data.  In particular, they are not able to simultaneously describe the polarization~\cite{LHCb:2013izl} and the \pt-differential cross section~\cite{LHCb:2011zfl}. A recent measurement by the LHCb Collaboration that looked at hadrons produced at small angles with respect to prompt \JPsi mesons (those not produced in {\PQb}-hadron decays) in proton-proton ($\Pp\Pp$) collisions~\cite{Aaij:2017fak} gives some new insight into this puzzle.  The observable is the \JPsi-jet fragmentation function, which corresponds to the distribution of $z$, the ratio of the \JPsi \pt to the \pt of the jet into which it is clustered. For jets in pseudorapidity $2.5 < \eta < 4.0$ and $\pt > 20\GeV$, LHCb observes that prompt \JPsi mesons are produced with far more in-jet associated hadroproduction than predicted by models, \ie, \JPsi mesons tend towards lower values of $z$.  Models of \JPsi production typically couple fixed-order perturbative quantum chromodynamics calculations with nonperturbative matrix elements that describe hadronization of the charm quark pair into a color-neutral state.  A solution to this discrepancy was proposed in Ref.~\cite{Bain:2017wvk}, where the evolution of the parton shower prior to the formation of the \JPsi is accounted for.  By including this parton shower contribution, which is not adequately described in hadronization generators, such as \PYTHIA~\cite{Sjostrand2015}, the authors were able to successfully describe the data.  Although the LHCb measurement only concerns the subset of \JPsi mesons found inside a relatively high-\pt jet, a recent measurement by CMS indicates that, for \JPsi mesons with energy larger than $15\GeV$,  nearly all of them are produced in association with a significant degree of jet activity~\cite{Sirunyan:2019vlp}. 

Assuming this explanation of the LHCb data is correct, this paradigm shift in our understanding of \JPsi production has important implications for the interpretation of \JPsi data in nucleus-nucleus collisions.  It implies that \JPsi are not exclusively produced at short times, but may also be produced in the course of the interaction of a hard-scattered parton with the QGP.  Hence, the suppression of the yield, as quantified by the nuclear modification factor (\RAA), may be sensitive to parton energy loss, the same phenomenon that gives rise to jet quenching.  There are already some hints in this direction.  First, as observed in Ref.~\cite{Arleo:2017ntr}, the \JPsi \RAA in \PbPb collisions~\cite{hin-16-025,Aaboud:2018quy} appears to exhibit the same rise with \pt as light hadrons show~\cite{Khachatryan:2016odn}, which for the case of light hadrons is well described by parton energy loss models~\cite{Chien:2015vja,Casalderrey-Solana:2014bpa,Bianchi:2017wpt,Xu:2015bbz,Andres:2016iys,Noronha-Hostler:2016eow}.  Second, in mid-central collisions the \JPsi show a significant magnitude of elliptic anisotropy in their azimuthal angle ($v_2$)~\cite{HIN-14-005,ATLAS:2018xms} at large \pt, a region where the hydrodynamical effects that produce such an anisotropy are expected to die out.  We know of no explanation for this high-\pt $v_2$ feature other than path-length-dependent parton energy loss. 

The goal of the current measurement is to investigate the $z$ dependence of the nuclear modification factor of jets containing a \JPsi meson, \ie,
\begin{equation}
R_{\mathrm{AA}}(z) = \frac{\rd N^{\tAA}/\rd z}{T_{\tAA} \rd\sigma^{\Pp\Pp}/\rd z}.
\end{equation}
This ratio compares the per-event yield in \PbPb collisions ($N^{\tAA}$) to the expectation based on $\Pp\Pp$ collisions, by scaling the cross section of the latter ($\sigma^{\Pp\Pp}$) by $T_{\tAA}$. The factor $T_{\tAA}$ is the average effective nucleon-nucleon luminosity delivered by a single heavy ion collision for a given centrality selection (a quantity related to the collision impact parameter)~\cite{Miller:2007ri}.  In the absence of nuclear effects, $R_{\tAA} = 1$.

This Letter constitutes the first direct study of the nuclear modification of \JPsi mesons inside jets.  Jets with a constituent \JPsi meson are studied in the jet \pt range of $30 < \pt < 40\GeV$ for 0--90\%, as well as 0--20\% and 20--90\% collision centralities.   The jets are required to be in the pseudorapidity range $\abs{\eta} < 2$, such that they are completely contained in the tracker acceptance, with no explicit selection on the rapidity of the \JPsi.   The \pt of the \JPsi is measured down to a threshold of 6.5\GeV.  This gives a range of $z$ from 0.22 to 1.  We investigate to what extent the \RAA varies with $z$, and indirectly, the formation time of the \JPsi.  These data potentially constrain the roles of the different QGP interaction mechanisms that may be at play, in particular parton energy loss and Debye screening.    

\section{The CMS detector}
\label{sec:detector}

The central feature of the CMS apparatus is a superconducting solenoid of 6\unit{m} internal diameter, providing a magnetic field of 3.8\unit{T}. Within the solenoid volume are a silicon pixel and strip tracker, a lead tungstate crystal electromagnetic calorimeter (ECAL), and a brass and scintillator hadron calorimeter (HCAL), each composed of a barrel and two endcap sections. Forward calorimeters extend the $\eta$ coverage provided by the barrel and endcap detectors. Muons are detected in gas-ionization chambers embedded in the steel flux-return yoke outside the solenoid.  Muons are measured in the range $\abs{\eta} < 2.4$, with detection planes made using three technologies: drift tubes, cathode strip chambers, and resistive plate chambers.  The efficiency to reconstruct and identify muons is greater than 96\% over the full range of $\eta$. Matching muons to tracks measured in the silicon tracker results in a relative transverse momentum resolution, for muons with \pt up to 100\GeV, of 1\% in the barrel and 3\% in the endcaps~\cite{Sirunyan:2018}. The forward hadron (HF) calorimeter uses steel as an absorber and quartz fibers as the sensitive material. The two halves of the HF are located 11.2\unit{m} from the interaction region, one on each end, and together they provide coverage in the range $3.0 < \abs{\eta} < 5.2$. 

Events of interest are selected using a two-tiered trigger system~\cite{Khachatryan:2016bia}. The first level (L1), composed of custom hardware processors, uses information from the calorimeters and muon detectors to select events at a rate of around 100\unit{kHz} for high luminosity $\Pp\Pp$ collisions and 30\unit{kHz} for \PbPb collisions.  The second level, known as the high-level trigger (HLT), consists of a farm of processors running a version of the full event reconstruction software optimized for fast processing, and reduces the event rate to around 1\unit{kHz} before data storage, for both $\Pp\Pp$ and \PbPb collisions.

The particle-flow algorithm~\cite{CMS-PRF-14-001} aims to reconstruct and identify each individual particle in an event, with an optimized combination of information from the various elements of the CMS detector. The energy of photons is obtained from the ECAL measurement. The energy of electrons is determined from a combination of the electron momentum at the primary interaction vertex as determined by the tracker, the energy of the corresponding ECAL cluster, and the energy sum of all bremsstrahlung photons spatially compatible with originating from the electron track. The energy of muons is obtained from the curvature of the corresponding track. The energy of charged hadrons is determined from a combination of their momentum measured in the tracker and the matching ECAL and HCAL energy deposits, corrected for the response function of the calorimeters to hadronic showers. Finally, the energy of neutral hadrons is obtained from the corresponding corrected ECAL and HCAL energies. 

Collision centrality is determined from the total transverse energy deposited in both of the HF calorimeters. The centrality is expressed as a percentage of the total inelastic hadronic cross section, with 0\% representing the most head-on (central) collisions and 100\% the most peripheral collisions. Hadronic events are selected by requiring at least three towers in each half of the HF calorimeter with an energy larger than 4\GeV. In this analysis, we restrict to the 90\% most central events, where the hadronic event selection is fully efficient.    

A more detailed description of the CMS detector, together with a definition of the coordinate system used and the relevant kinematic variables, can be found in Ref.~\cite{Chatrchyan:2008zzk}. 

\section{Data analysis}
\subsection{Simulation}

Monte Carlo simulations are used as a baseline for the \JPsi efficiency and acceptance, as well as the detector response to jets.  Correction factors to account for data-to-simulation discrepancies are discussed in the following section.  Events are generated with \PYTHIA 8 (version 8.230)~\cite{Sjostrand2015}, using the CP5 underlying event tune~\cite{Sirunyan:2019dfx}.  Prompt \JPsi are produced using all color-singlet and color-octet modes with the default matrix elements~\cite{Nason:1999ta}.  Simulation of \PbPb collisions is done by embedding \PYTHIA 8 events into \PbPb collision events produced with the \HYDJET generator (version 1.9)~\cite{Lokhtin:2005px}.  The event activity at mid-rapidity is matched to that of minimum bias collision data via an analysis of the energy deposited in randomly distributed cones.  The response of the CMS detector is simulated using the \GEANTfour package~\cite{Agostinelli:2002hh}.

\subsection{Prompt \texorpdfstring{\JPsi}{J/psi} meson signal extraction}

\JPsi mesons are measured via their decays into oppositely charged muon pairs.  Events are selected with a trigger that requires that at least two muon candidates are reconstructed in the muon subsystem, first at L1, and then using refined information at HLT.  In \PbPb collisions, the rate is reduced at HLT by requiring one of the two muons to match a track from the silicon tracking subsystem.  The invariant mass of the pair ($m_{\PGm\PGm}$) is required to be within the range of 1 to 5\GeV.

\begin{figure}[hbt!]
  \begin{center}
\includegraphics[width=0.48\textwidth]{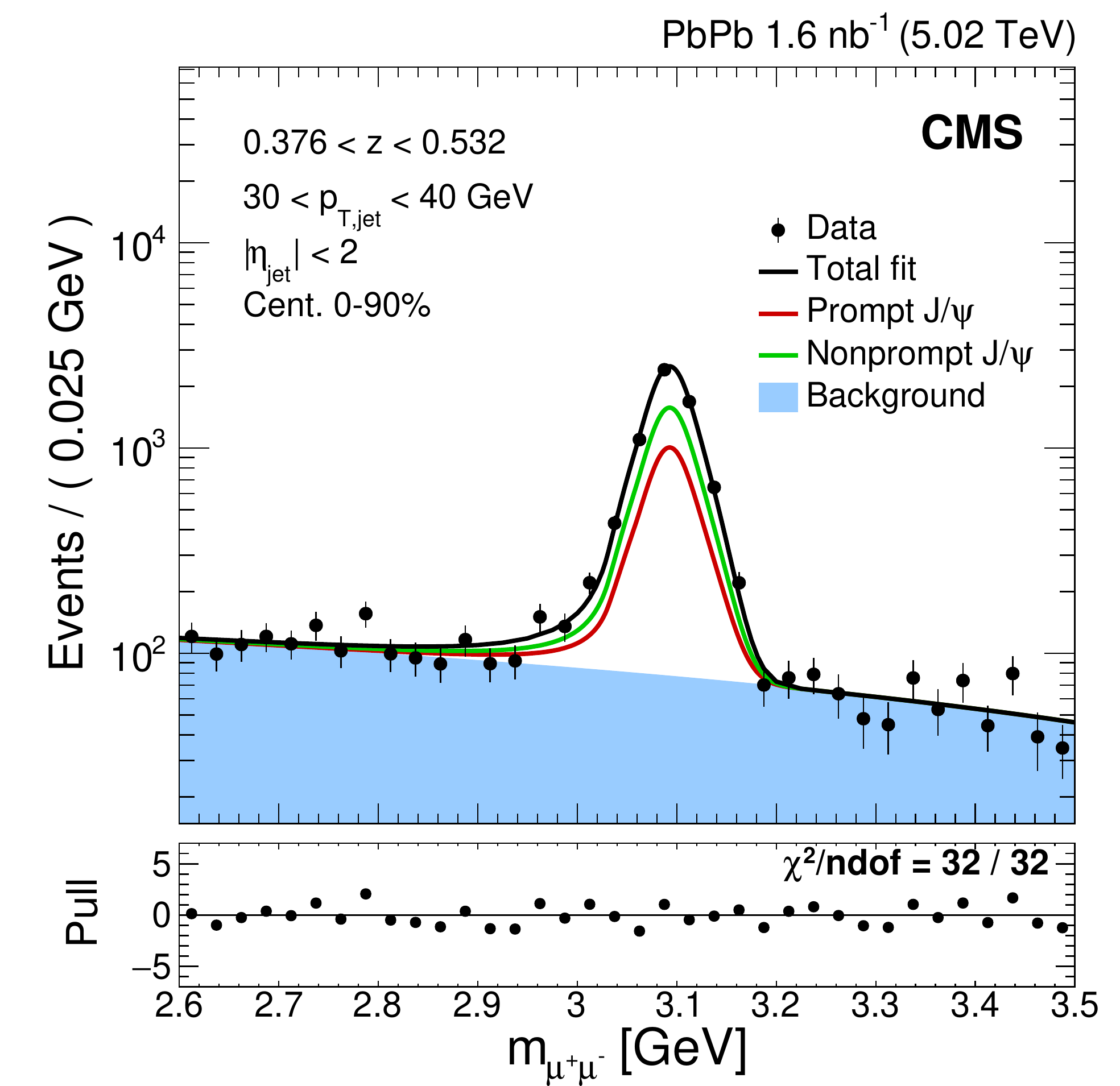}
\includegraphics[width=0.48\textwidth]{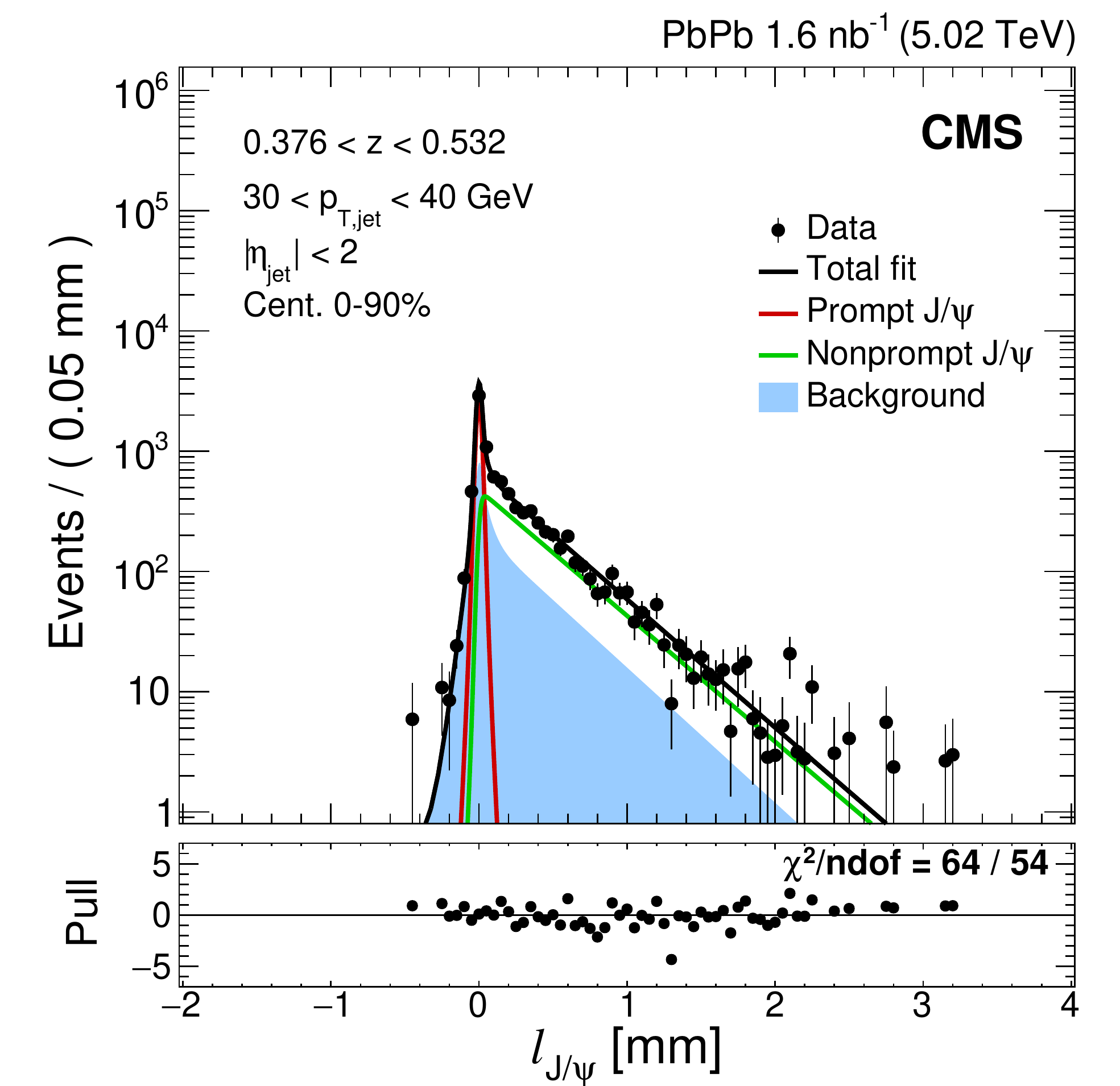}
  \caption{Projections of a two-dimensional fit used to extract the prompt \JPsi yield in \PbPb collisions.  \cmsLeftI:  The dimuon invariant mass distribution.  \cmsRightI:  The pseudo-proper decay length ($l_{\JPsi}$) distribution.  The fit functions are described in the text. The lower panels display the pull distributions, defined as the difference between the fit and the data, divided by the combined statistical uncertainty in the fit and in the data.}
    \label{fig:fits}
  \end{center}
\end{figure}

A set of offline muon selection criteria that is optimized for low \pt muons is applied~\cite{hin-16-025}.  The \JPsi yield is obtained from a fit to the $m_{\PGm\PGm}$ distribution in bins of jet \pt and $z$.  An example of a $m_{\PGm\PGm}$ fit is shown in the \cmsLeft panel of Fig.~\ref{fig:fits}.  The signal is modeled as the sum of two Crystal Ball functions~\cite{Oreglia:1980cs} with different widths, but common mean and tail parameters.  The tail parameters are obtained from simulation.  The dimuon background is modeled by a Chebyshev polynomial.

Nonprompt \JPsi mesons, \ie, those that are produced in \PQb hadron decays, are separated from the prompt component by exploiting the long lifetime of these decays.  A fit is performed to the distribution of the pseudo-proper decay length, $l_{\JPsi} = L m / \abs{p}$, where $L$ is the distance along the beam axis between the \JPsi vertex and the nearest primary collision vertex, and $m$ and $p$ are the world-average \JPsi mass (3.097\GeV)~\cite{Tanabashi:2018oca} and the \JPsi candidate momentum, respectively.  An example of a $l_{\JPsi}$ fit is shown in the \cmsRight panel of Fig.~\ref{fig:fits}.  In order to parameterize the $l_{\JPsi}$ distribution, the sPlot technique~\cite{Pivk:2004ty} is used to decompose it into the \JPsi signal and non-\JPsi background components, using $m_{\PGm\PGm}$ as the discriminating variable.  Prompt \JPsi mesons can have nonzero values of $l_{\JPsi}$, both positive and negative, because of the finite detector resolution.  The resolution is modeled as a sum of two Gaussian functions, which are constrained by fitting the negative part of the $l_{\JPsi}$ distribution. The nonprompt component is modeled as an exponential function, with the lifetime treated as a free parameter, convolved with the same double-Gaussian resolution function.  The combinatorial background is itself well described by prompt- and nonprompt-like components, again using the same resolution function.   The nonprompt component of the background is described by an empirical sum of exponential functions.  To extract the yield of prompt \JPsi mesons a two-dimensional fit to the joint $m_{\PGm\PGm}$ and $l_{\JPsi}$ distribution is performed.  All parameters are fixed by the one-dimensional fits, aside from the prompt and nonprompt \JPsi signal normalizations, and the background normalizations, as well as the nonprompt signal lifetime parameter.  A more detailed description of the fitting procedure can be found in Ref.~\cite{hin-16-025}.

The \JPsi meson acceptance, and the efficiency of reconstruction and identification, are determined in simulation in finely binned histograms of the \pt and $\eta$ of the \JPsi.  For \PbPb collisions, the efficiency is also determined in bins of collision centrality.  The correction for acceptance and efficiency is applied as a weight factor to each \JPsi candidate prior to the signal extraction.  Differences between data and simulation are evaluated \textit{in-situ}, from events collected with a single-muon trigger, using the tag-and-probe technique~\cite{Chatrchyan:2012xi}.  These additional corrections are applied to the simulation in relatively coarse bins.  Tag-and-probe corrections are derived separately for trigger, identification, and reconstruction efficiency.

\subsection{Jet \texorpdfstring{\pt}{pt} determination}

Jets are clustered using the anti-\kt algorithm~\cite{Cacciari:2008gp,Cacciari:2011ma} with a distance parameter of $R = 0.3$.  The constituents of the jets are the output of the particle-flow algorithm, described in Section~\ref{sec:detector}, with the exception of the \JPsi candidate.  Muon pairs with an invariant mass consistent with a \JPsi decay are replaced by the reconstructed \JPsi candidate.   Jet energy corrections are derived from simulation as a function of \pt and $\eta$ using the framework described in Ref.~\cite{Khachatryan:2016kdb}, in which data-to-simulation scale factors are obtained from \PZ{}+jet, \PGg{}+jet and dijet \pt balancing studies.  In the case of \PbPb collisions, these corrections are derived from peripheral events to avoid additional \pt imbalance from jet quenching.  The jet energy correction procedure is only applied to the non-\JPsi component of the jet, as the momentum of the \JPsi is determined with high precision from its dimuon decay.  In \PbPb collisions, the constituent subtraction method~\cite{Berta:2014eza} is employed to subtract the contribution to the jet momentum from the underlying event.  As the signal particle of interest, the \JPsi meson is excluded from the subtraction, as it would be incorrect to attribute any fraction of its energy to the underlying event. 

In addition to the energy scale, the energy resolution of jets is somewhat degraded in data compared to simulation.  The corresponding scale factors are derived from dijet balancing studies performed in $\Pp\Pp$ collisions at $\sqrt{s} = 13\TeV$ in 2017 and 2018~\cite{Khachatryan:2016kdb}.  The resolution is found to be between 10 and 20\% worse than the simulation, depending on $\eta$.  In order to incorporate this effect, a Gaussian smearing is applied to the measured jet \pt values in simulation to match the resolution in data.  This smeared mapping is then applied in the unfolding procedure, as described below. 

\begin{figure}[hbt!]
  \begin{center}
\includegraphics[width=0.48\textwidth]{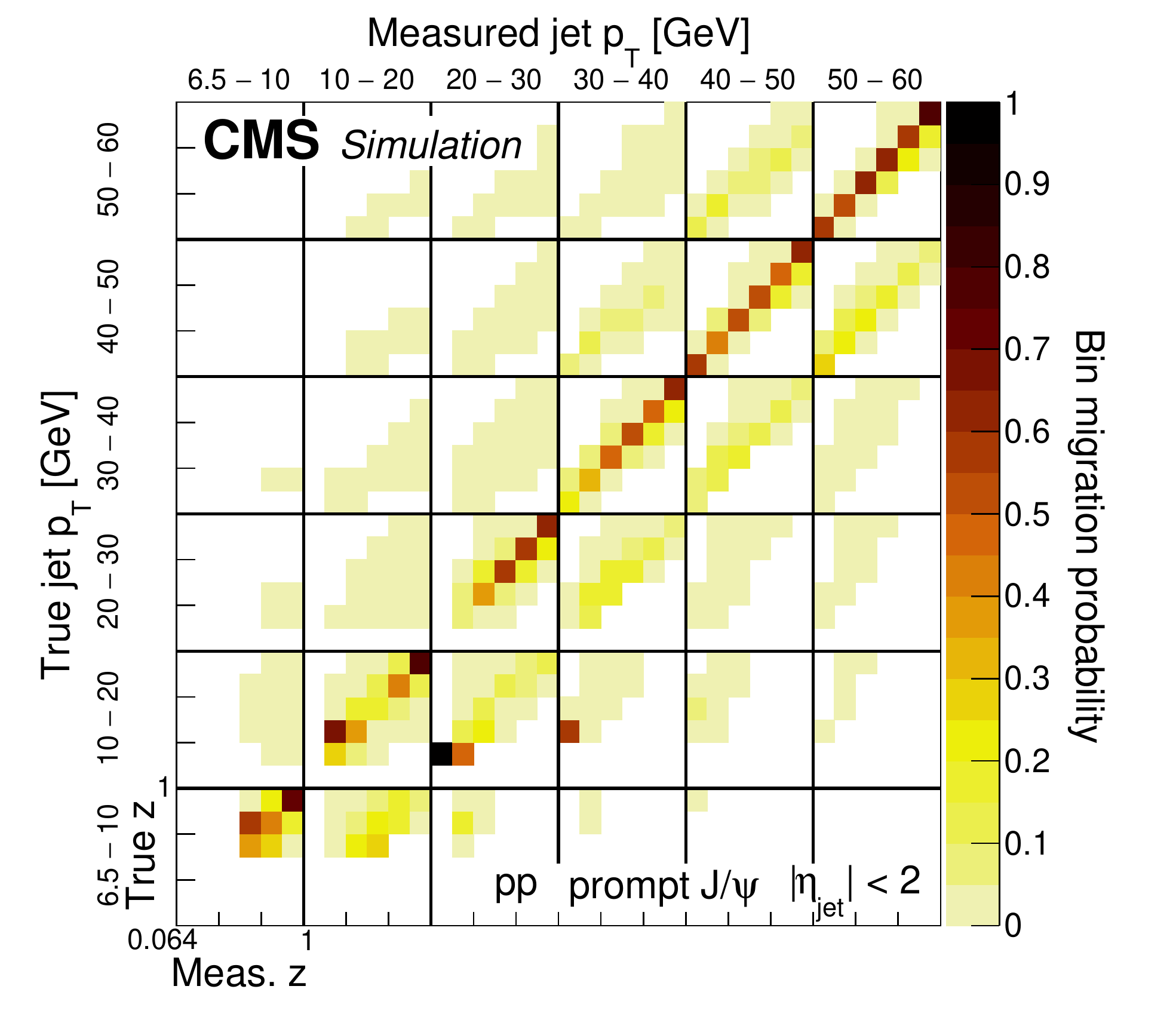}
\includegraphics[width=0.48\textwidth]{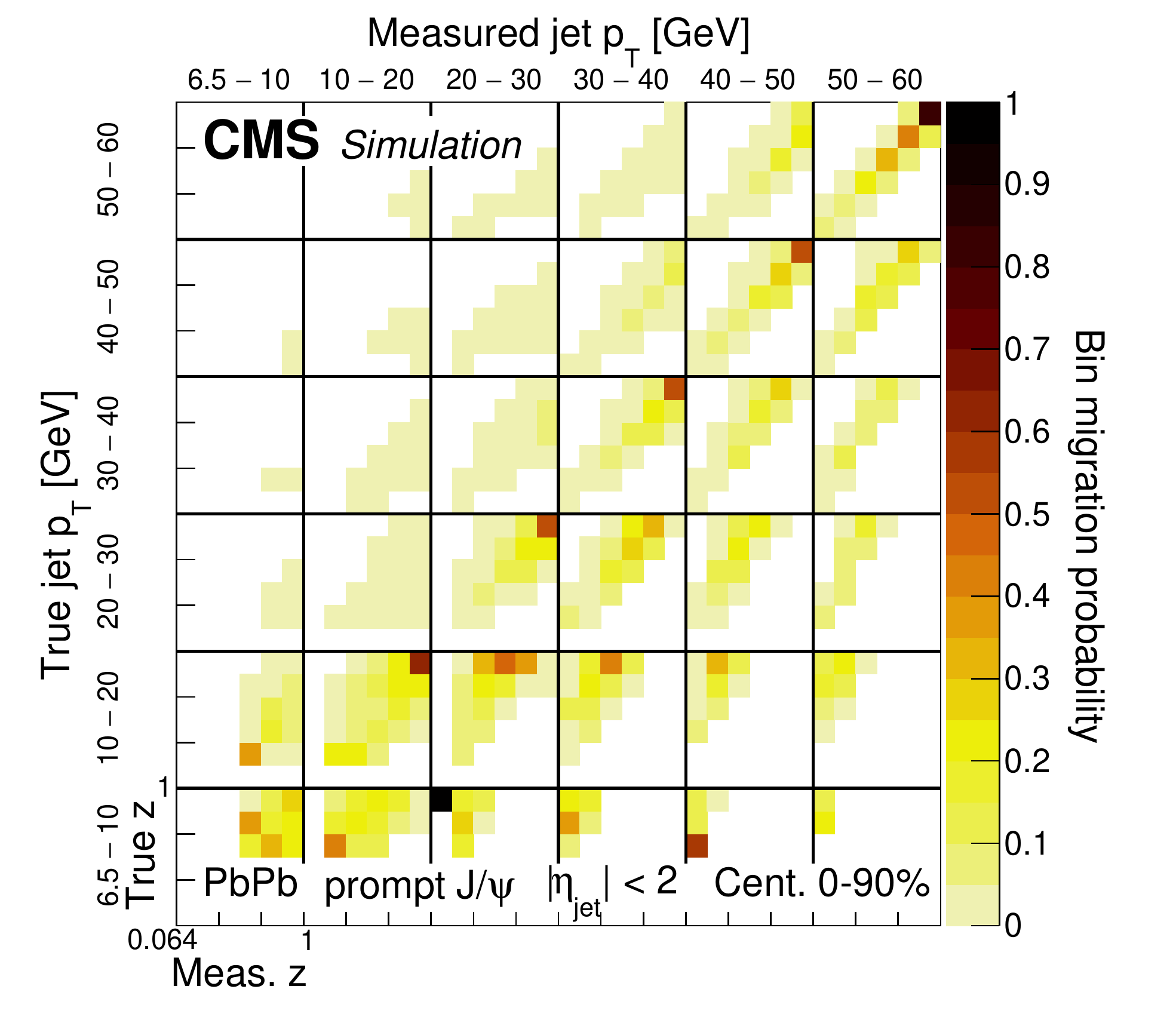}
  \caption{
   Detector response matrices for jets containing a prompt \JPsi meson, showing the bin migration probability as a function of jet \pt and~$z$.  The response for $\Pp\Pp$ collisions (\cmsLeft) is evaluated using \PYTHIA 8.  The response for \PbPb collisions (\cmsRight) is evaluated using \PYTHIA 8 embedded into \HYDJET.
  }
    \label{fig:trMatrix}
  \end{center}
\end{figure}

The finite \pt resolution of jets causes migration of jets across $z$ bins.  It also causes migration of jets into and out of the nominal \pt selection of 30--40\GeV.  In order to capture the full jet response, we also perform the measurement in underflow and overflow bins of 10--30\GeV and 40--50\GeV, respectively.  With the current data, a reliable yield cannot be extracted for \pt intervals larger than the selected overflow bin, which motivates the choice of the nominal \pt interval.  To correct for these bin migration effects, we unfold the data simultaneously in these two dimensions (jet \pt and $z$) using the iterative D'Agostini method~\cite{DAgostini:1994fjx}, as implemented in the \textsc{RooUnfold} package~\cite{Adye:1349242}.  The detector response matrix, which defines the relationship between the true and measured values, is taken from the \PYTHIA 8 simulation, embedded into \HYDJET for the case of \PbPb collisions, aside from the aforementioned data-to-simulation scale factors for the jet energy scale and resolution.   The response matrices for $\Pp\Pp$ and \PbPb collisions are shown in Fig.~\ref{fig:trMatrix}.  The coarse bins show the jet \pt dependence, whereas the finer inset bins show the dependence on $z$. Each column of measured bins is normalized to unity, such that each bin represents the fractional contribution of the given true \pt and $z$ values to the measured values for that column.  The \PbPb data exhibit substantially larger off-diagonal contributions than for $\Pp\Pp$, as expected from the larger underlying event, which drives the bin migration in \PbPb.

In this method, the unfolding is initialized with a ``prior'' distribution in the two variables, which is taken from simulation.  To avoid bias from the presumed shape of the $z$ distribution in simulation, which is known to be inaccurate~\cite{Aaij:2017fak}, the prior distribution is flattened in $z$.  After a tunable number of iterations, which corresponds to the degree of regularization, the procedure is truncated.  To improve the performance of the unfolding, we run the full set of iterations multiple times.  The prior of each such ``super-iteration'' is obtained using the $z$ distribution that is output from the previous super-iteration.  The numbers of iterations and super-iterations are optimized using simulated prompt \JPsi events, by applying a random Gaussian smearing to the measured $z$ distributions according to the relative statistical uncertainties in data, in order to emulate the effect of statistical fluctuations.  The optimization criterion is the minimum $\chi^2$ between the unfolded and true $z$ distributions. The use of three iterations is found to give the best performance in both $\Pp\Pp$ and \PbPb simulated samples.  However, while only one super-iteration is sufficient for $\Pp\Pp$, 25 super-iterations are required for \PbPb collisions.  Since the optimal settings depend not only on the statistical uncertainties of the measured $z$ distribution, but also on its shape, we additionally evaluated them in nonprompt \JPsi meson simulation, where the measured distribution is more similar to the data than in the prompt \JPsi meson simulation, by unfolding the nonprompt $z$ distributions with the nonprompt response matrices and using the settings with the best performance as a systematic variation in the regularization procedure.

\section{Systematic uncertainties}

The systematic uncertainties may be divided into three categories:  \JPsi signal extraction, jet energy scale and resolution, and normalization uncertainties.

\textit{\JPsi signal extraction}:  Uncertainties in the extraction of the \JPsi yields arise from the signal and background shape modeling, as well as from the acceptance and efficiency of muon reconstruction and identification.  The signal shape of the $m_{\PGm\PGm}$ distribution is varied from the two Crystal Ball parameterization to a single Crystal Ball function convoluted with a Gaussian function.  The radiative tail of the signal model is also varied by treating the tail parameters of the Crystal Ball function as free parameters, rather than taking their values from simulation.  The relative uncertainty in the \JPsi yield coming from the signal modeling is less than 1\% in $\Pp\Pp$ and less than 2\% in \PbPb collisions.  The dependence of the signal shape on $l_{\JPsi}$ is evaluated by switching from a parameterization of the lifetime distribution of nonprompt \JPsi signal to a template built from simulation.  The relative uncertainty from this variation is less than 1.5\% in both $\Pp\Pp$ and \PbPb collisions, except at low \JPsi \pt ($<$10\GeV) in $\Pp\Pp$ collisions, where it reaches 4.5\%, and where it dominates the uncertainties in the \JPsi signal extraction.  The Gaussian function describing the lifetime resolution is varied to use the value obtained from simulation, rather than obtaining it from the data.  This variation affects both the signal and background modeling, and results in a relative uncertainty of less than 2\% in $\Pp\Pp$ and \PbPb collisions. 

The uncertainty in the description of the background is obtained as follows.  Instead of a Chebyshev polynomial, an exponential of a Chebyshev polynomial is used to describe the $m_{\PGm\PGm}$ distribution.  The range of $m_{\PGm\PGm}$ used in the fit is varied as well, to reduce the contamination from the $\psi$(2S) state.  This results in an uncertainty of less than 1\% in $\Pp\Pp$ and \PbPb collisions, except at low \JPsi \pt ($< 10$\GeV) in \PbPb collisions, where it reaches 6\% and dominates the uncertainties in the \JPsi signal extraction. As was done for the signal, a template is used to describe the $l_{\JPsi}$ distribution, rather than a parameterization.  In this case, the template is obtained from a background-like event sample using sPlot.  The source results in a relative uncertainty of $< 2$\% in the \JPsi yields in both $\Pp\Pp$ and \PbPb collisions. 

The acceptance and efficiency corrections lead to an associated relative uncertainty in the \JPsi yields in $\Pp\Pp$ collisions of 1.5--2.5\%, and 3--4\% in \PbPb collisions, depending on the value of \JPsi \pt, and comprise the dominant source of uncertainty in the \JPsi signal extraction for \JPsi ${\pt  > 10\GeV}$.  The largest component of the uncertainty is statistical in nature, coming from the finite size of the single-muon trigger samples used in the tag-and-probe method.  The statistical uncertainty of the simulation samples is also taken into account, but is generally subdominant.  A systematic component is evaluated by variations of the signal and background modeling in the extraction of the \JPsi yield from fits to the  $m_{\PGm\PGm}$ distribution, similar to the procedure for the \JPsi yield extraction in the main analysis, described above.  The relative uncertainty from this source is around 1\% in both $\Pp\Pp$ and \PbPb collisions, with no strong dependence on \pt.  Further details on the procedure used to evaluate systematic uncertainties for muons from the tag-and-probe method are reported in Refs.~\cite{Chatrchyan:2012xi,Sirunyan:2018}.

\textit{Jet energy scale and resolution}:  The uncertainty in the jet energy scale is evaluated from dijet and \PGg{}+jet balancing methods, as described in Ref.~\cite{Khachatryan:2016kdb}.  The uncertainty in the jet energy scale in $\Pp\Pp$ collisions is around 3--4\%, increasing as a function of $\abs{\eta}$.  In \PbPb collisions, the uncertainty is around 4\%, except for the barrel-endcap transition region ($1.3 < \abs{\eta} < 1.6$), where it can become as large as 10\%. An additional source of uncertainty in the jet energy scale is considered in \PbPb collisions to take into account the differences in the particle mixture in simulation and data due to jet quenching, as described in Ref.~\cite{CMS:2017ehl}. This uncertainty is around 5\% for the 30\% most central events and 2.5\% for all other centralities.  The uncertainty in the jet energy resolution is evaluated from a dijet balancing method, as also described in Ref.~\cite{Khachatryan:2016kdb}. In $\Pp\Pp$ collisions, the uncertainty in the resolution is in the range of 2--4\% in the barrel region, but is larger in the endcap and transition regions, where it varies in the range of 10--20\%, depending on $\eta$.  For \PbPb collisions, the uncertainty is evaluated from peripheral data, as well as from $\Pp\Pp$ data recorded in the same yearly running period, and varies from 3\% in the barrel to 7\% in the endcap and transition regions.  In \PbPb, there is an additional contribution to the uncertainty because of the modeling of the underlying event in simulation with \HYDJET.  This uncertainty is evaluated by comparing the energy in randomly distributed cones in data and simulation.  The difference between the random cone distributions in data and simulation is used to estimate the effect on the jet resolution.

\textit{Unfolding}:  The uncertainty in the regularization applied is estimated by varying the number of iterations used in the unfolding, from the settings found to be optimal for prompt \JPsi to those found to be optimal for the nonprompt \JPsi signal, which has a very different signal shape as a function of $z$. For nonprompt simulations in $\Pp\Pp$ and \PbPb collisions, the use of ten iterations is found to give the best performance, in contrast to three iterations for the prompt simulation.  In both cases, the statistical precision of the data is emulated in simulation.  The assumption of a prior that is flat in $z$ is also relaxed, by instead initializing the prior to match the shape of the truth $z$ distributions in simulated nonprompt \JPsi signal samples,  which are more similar in shape to the measured $z$ distributions.   Finally, the statistical uncertainty of the simulated samples used in the unfolding is taken into account, and is generally subdominant.   Systematic uncertainties related to the jet energy scale and resolution are propagated by producing variations of the detector response matrix and repeating the unfolding procedure for each variation.

\textit{Normalization uncertainties}:  Cross section measurements in $\Pp\Pp$ collisions have an overall uncertainty from the integrated luminosity of 1.9\%~\cite{CMS-PAS-LUM-19-001} that is obtained from an analysis of data from a van der Meer scan~\cite{Meer1968}.  The \PbPb data are normalized by the equivalent number of hadronic nucleon-nucleon interactions in the data sample, which has an uncertainty of 1.3\% coming from the selection of such events, taking into account possible contamination from electromagnetic interactions and beam backgrounds.  To compare to data from $\Pp\Pp$ collisions, the \PbPb data are normalized by $T_{\tAA}$, which is determined from the Monte Carlo implementation of the Glauber model described in Ref.~\cite{Loizides:2017ack}.  For the centrality intervals used in this analysis, the corresponding values of $T_{\tAA}$ are $6.27 \pm 0.14$, $18.79 \pm 0.36$, and $2.717 \pm 0.098 \mbinv$, for the 0--90, 0--20 and 20--90\% centrality intervals, respectively.

\begin{figure}[hbt!]
  \begin{center}
\includegraphics[width=0.48\textwidth]{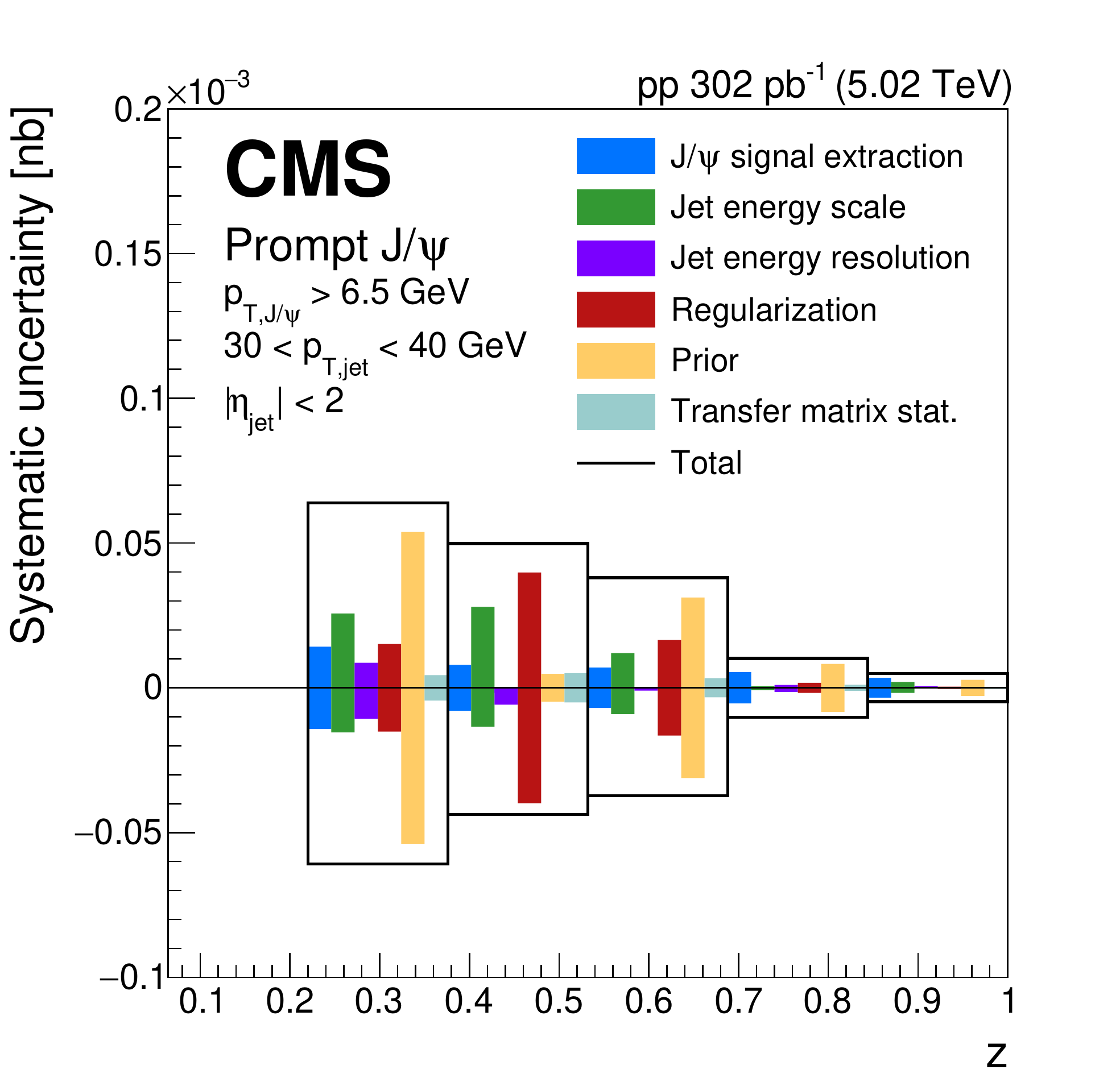}
\includegraphics[width=0.48\textwidth]{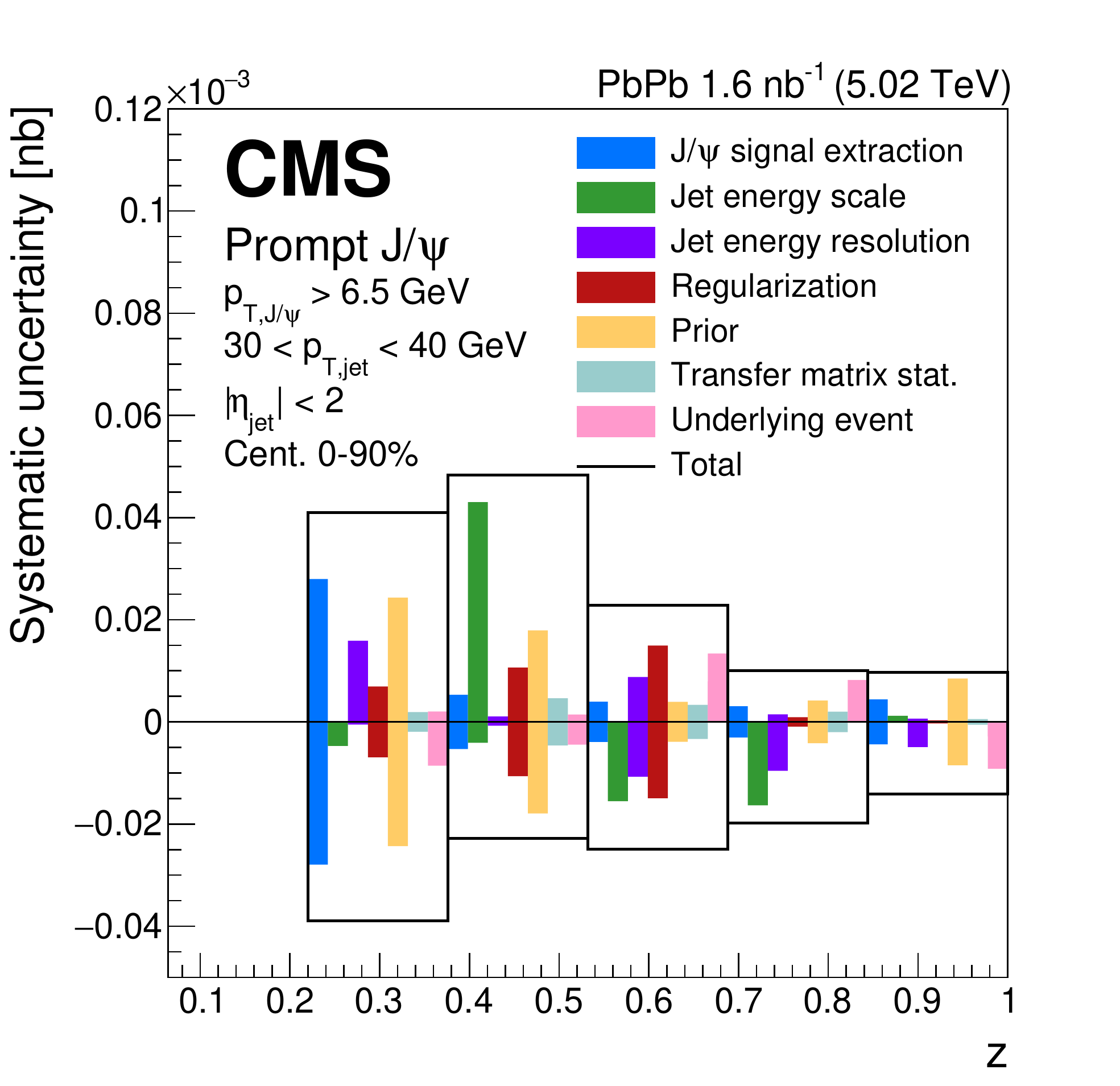}
  \caption{The main sources of systematic uncertainty, as a function of $z$, for the cross section in $\Pp\Pp$ (\cmsLeft) and the $T_{\tAA}$-scaled yield in \PbPb (\cmsRight) collisions. The normalization uncertainties of 1.9\% for $\Pp\Pp$ and 2.6\% for \PbPb are not shown. 
}
    \label{fig:systematics}
  \end{center}
\end{figure}

The various sources of systematic uncertainty are shown in Fig.~\ref{fig:systematics}.  Sources related to the \JPsi yield extraction are combined into a single component, for visibility.  The total uncertainty is the quadrature sum of all the sources.  Some sources of systematic uncertainty are largely correlated bin-to-bin, notably the uncertainty in the jet energy scale, jet energy resolution, and prior.  No cancellation of systematic uncertainties is assumed, however, when comparing $\Pp\Pp$ to \PbPb collisions.

\section{Results}

\begin{figure}[hbtp]
  \begin{center}
\includegraphics[width=0.48\textwidth]{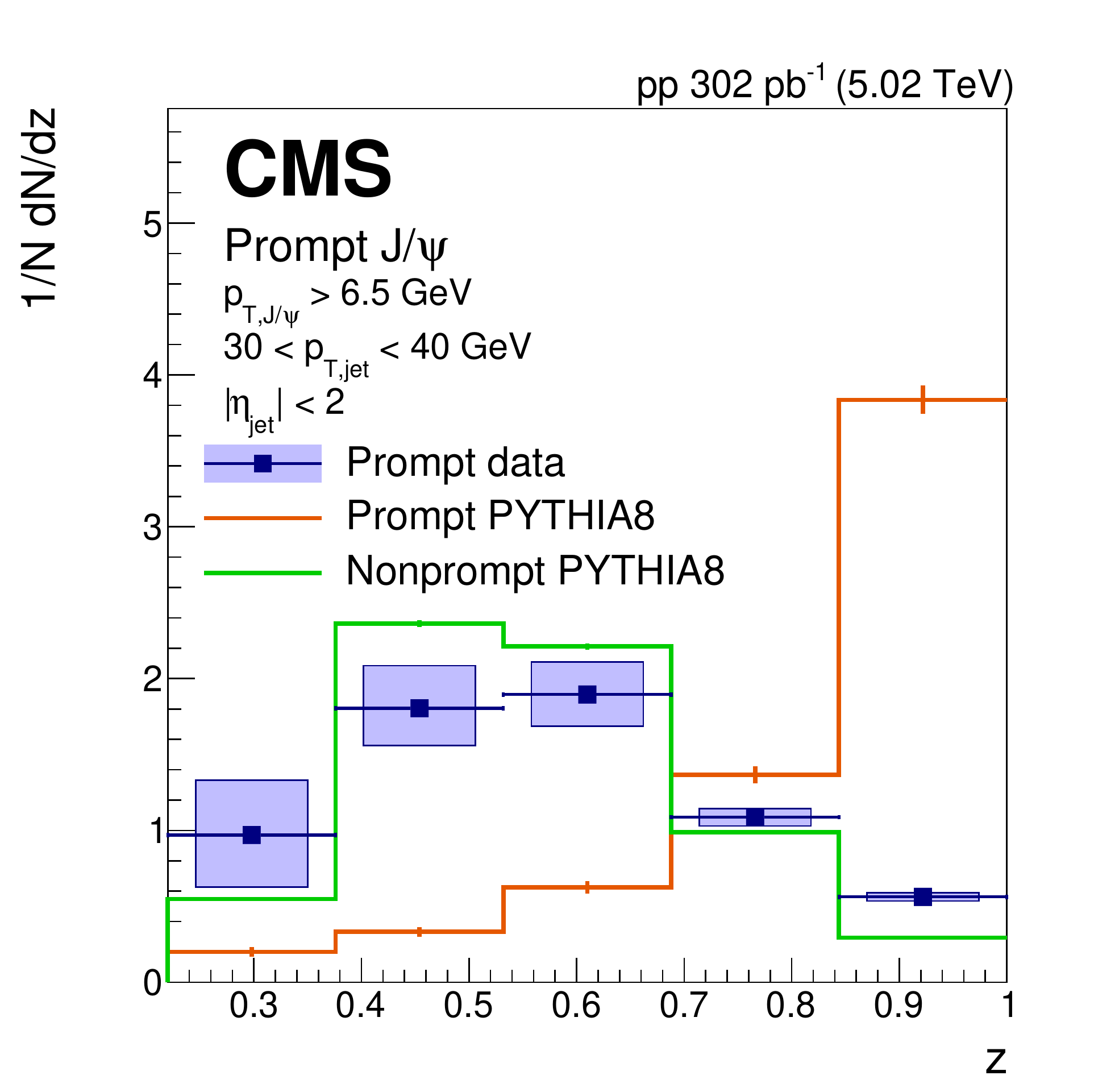}
  \caption{Normalized $z$ distribution in $\Pp\Pp$ collisions, compared to prompt and nonprompt \JPsi in \PYTHIA 8, at the generator level.  Bars indicate statistical uncertainties, while systematic uncertainties are depicted as boxes.}
    \label{fig:zDistPP}
  \end{center}
\end{figure}

Figure~\ref{fig:zDistPP} shows the distribution in $\Pp\Pp$ data of the fragmentation variable $z$ for prompt \JPsi mesons.  Its shape is compared to generator-level predictions from \PYTHIA 8 for prompt and nonprompt \JPsi signals.  In contrast to the \PYTHIA 8 simulation, where prompt \JPsi are produced directly in the matrix element partonic scattering, the data show a relatively large degree of surrounding jet activity, indicative of \JPsi production inside of parton showers.   The $z$ distribution in data more closely resembles that of the nonprompt \JPsi\ \PYTHIA 8 simulation, which contains a larger jet-like component from fragmentation, as well as other products of the {\PQb}-hadron decay.  The data confirm the trends observed in Ref.~\cite{Aaij:2017fak}, but in a different rapidity range and nucleon-nucleon center-of-mass energy.

\begin{figure}[hbt!]
  \begin{center}
\includegraphics[width=0.48\textwidth]{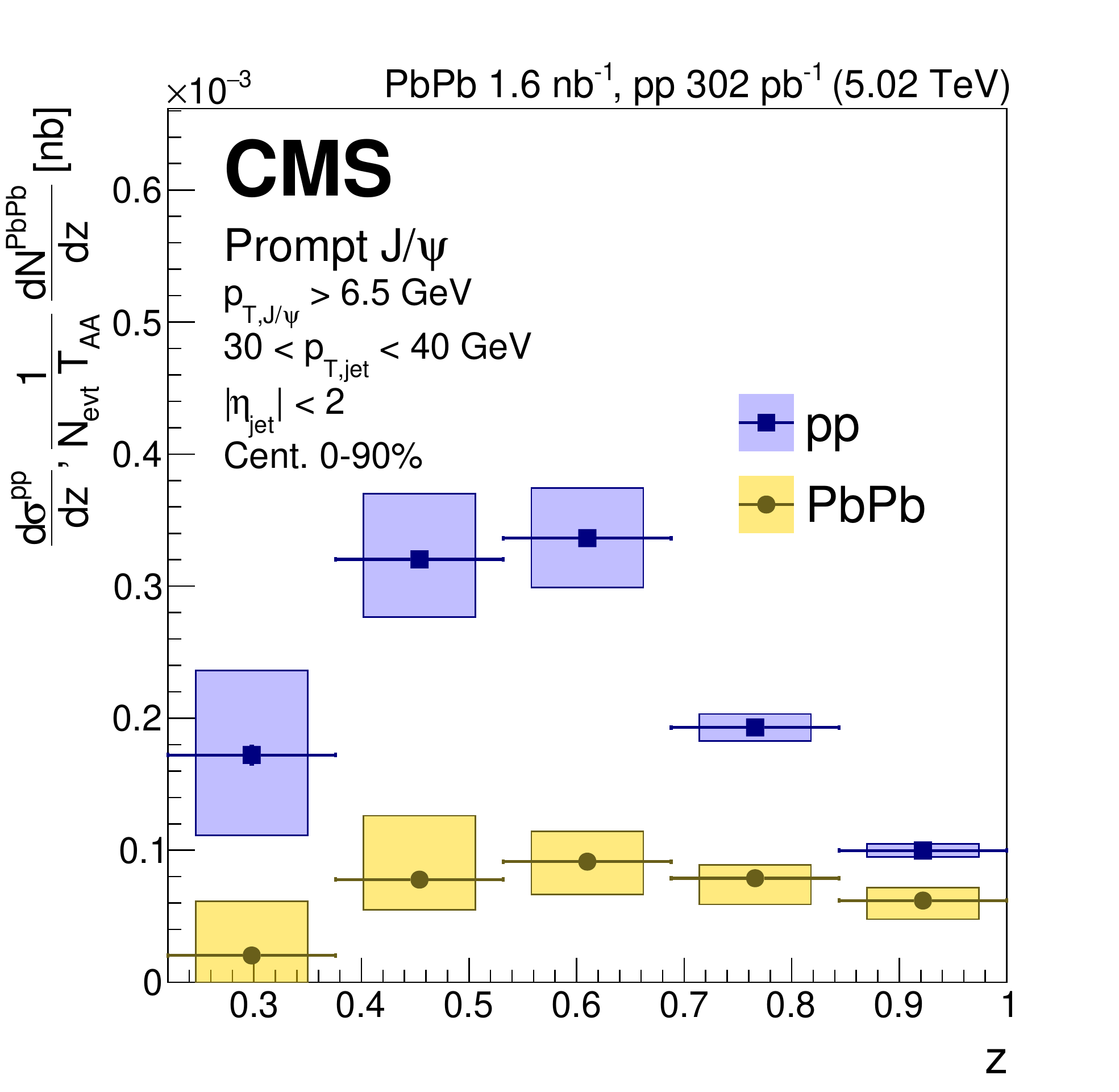}
\includegraphics[width=0.48\textwidth]{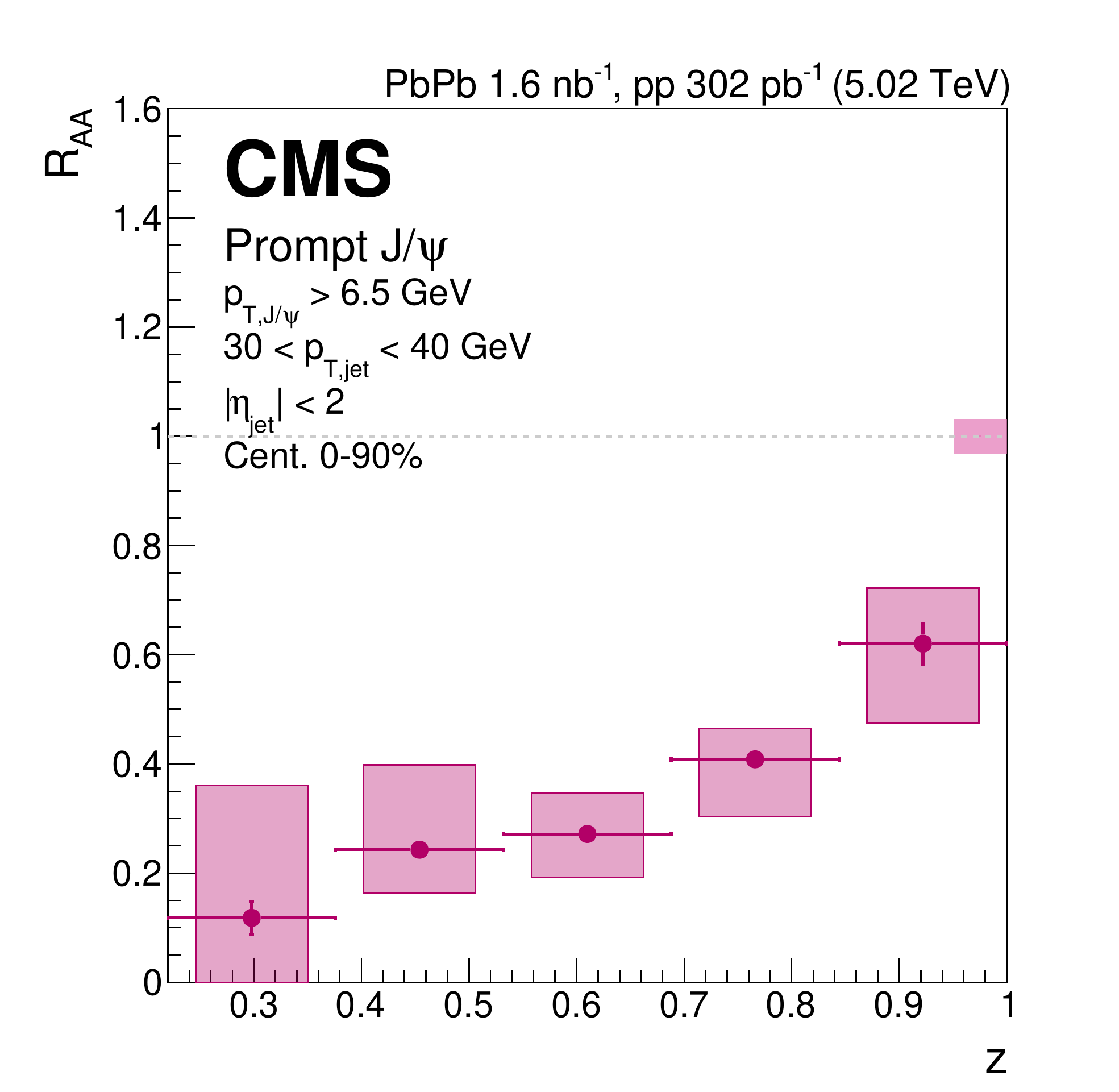}
  \caption{\cmsLeftI:  The \JPsi differential cross section in $\Pp\Pp$ and the $T_{\tAA}$-scaled yield in \PbPb collisions, as a function of $z$. The normalization uncertainties of 1.9\% for $\Pp\Pp$ and 2.6\% for \PbPb are not included in the  point-by-point uncertainty boxes and not shown in the plot.  \cmsRightI:  The nuclear modification factor \RAA, as a function of $z$.  Bars indicate statistical uncertainties, while systematic uncertainties are depicted as boxes.  The box around unity shows the normalization uncertainty. }
    \label{fig:results}
  \end{center}
\end{figure}

Figure~\ref{fig:results} (\cmsLeft) shows the same $z$ distribution in $\Pp\Pp$ collisions, this time normalized as a differential cross section.  The per-event yield of prompt \JPsi mesons in \PbPb collisions is also shown.   In order to compare the two collision systems, the \PbPb yields are scaled by the nuclear overlap factor $T_{\tAA}$.  The \PbPb data are also peaked at an intermediate value of $z$, indicating a sizable amount of jet activity. 

\begin{figure}[hbtp]
  \begin{center}
\includegraphics[width=0.48\textwidth]{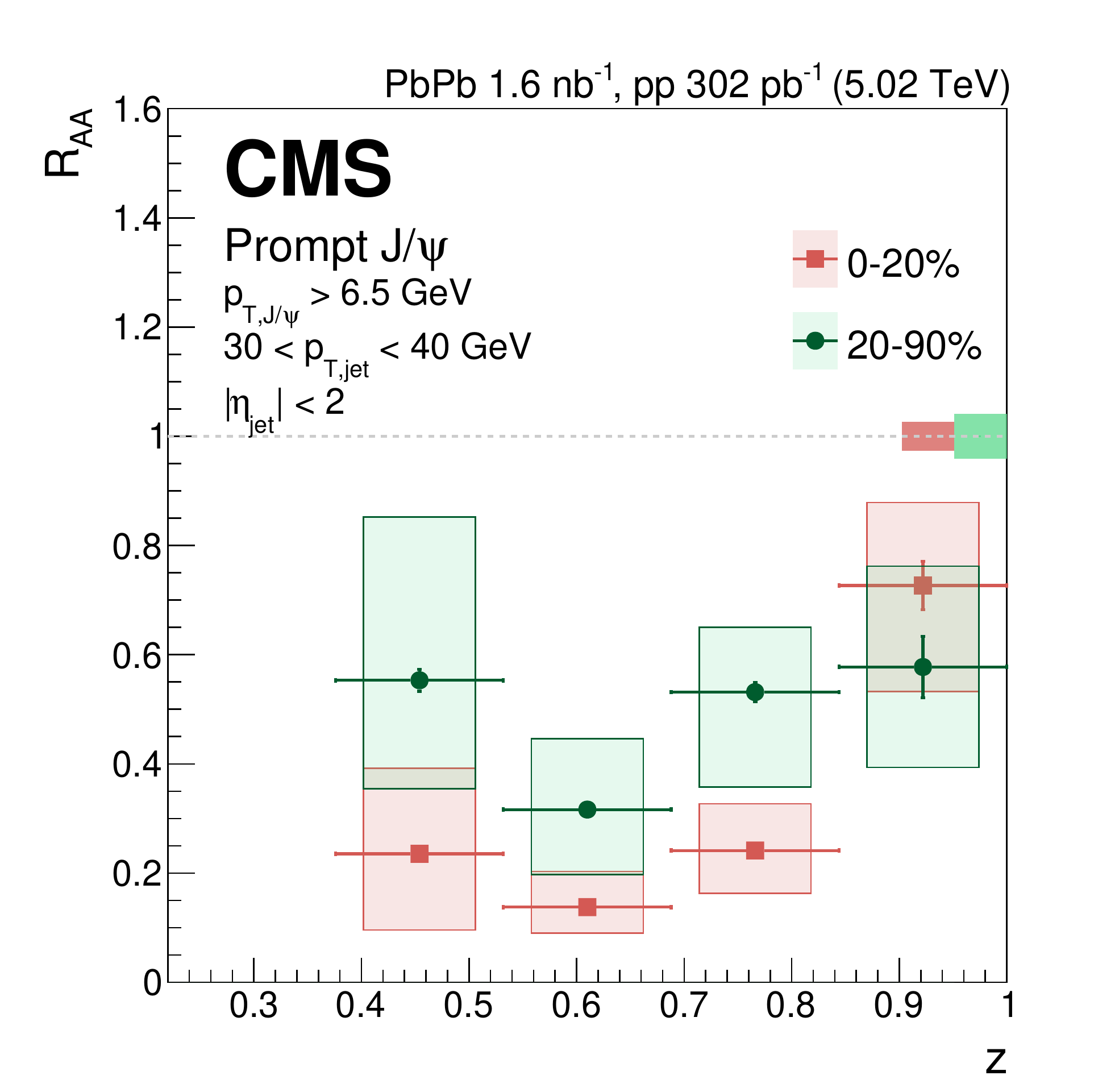}
  \caption{The nuclear modification factor \RAA for two centrality selections of \PbPb collisions, as a function of $z$. Due to limited statistical precision, the lowest $z$ bin is excluded. Bars indicate statistical uncertainties, while systematic uncertainties are depicted as boxes.  The boxes around unity show the normalization uncertainties. }
    \label{fig:resultsCent}
  \end{center}
\end{figure}

The \PbPb data show a suppression level that is generally comparable to that observed for ``inclusive'' prompt \JPsi production, \ie, without an explicit jet requirement~\cite{hin-16-025}.   This is quantified by the ratio of these two distributions, \RAA, shown in Fig.~\ref{fig:results} (\cmsRight).  The data show a slight rising trend as a function of $z$, with a significance of around two standard deviations.   Fig.~\ref{fig:resultsCent} shows the \RAA for two centrality selections, 0--20 and 20--90\%.  A larger degree of suppression is suggested for the more central selection, as expected for final-state effects related to the QGP.   The rising trend with increasing $z$ is somewhat more pronounced in central events.

In the largest $z$ bin, where the \JPsi is produced with fewer associated particles, the suppression is significantly reduced as compared to lower values of $z$, most importantly in the centrality-integrated results.  Such a reduction of suppression at large $z$ has a natural interpretation in terms of the jet quenching phenomenon.   Lower values of $z$ should be populated with jets with a \JPsi produced late in the parton shower.  Such a parton cascade is expected to have a large degree of interaction with the QGP in the form of subsequent medium-induced emissions, as compared to a jet with a small partonic multiplicity~\cite{Caucal:2018dla}.  In this picture, the rising trend observed for inclusive prompt \JPsi production would be explained by the same mechanism, as $z$ tends to increase with increasing \pt.

\section{Summary}

Jets containing a prompt \JPsi meson were studied in proton-proton ($\Pp\Pp$) and lead-lead (\PbPb) collisions at $\sqrtsNN = 5.02\TeV$, for jets with transverse momentum $30 < \pt < 40\GeV$ and pseudorapidity $\abs{\eta} < 2$.  The distribution of the fragmentation variable $z$, the ratio of the \JPsi \pt to that of the jet, was measured in both systems.  In $\Pp\Pp$ collisions, prompt \JPsi mesons were found to have more surrounding jet activity, \ie, to populate lower values of $z$ than predicted by \PYTHIA 8 simulations, suggesting that \JPsi production late in the parton shower is underestimated.  The $\Pp\Pp$ and \PbPb distributions were compared by calculating the nuclear modification factor, \RAA, the ratio of \PbPb data to the expectation based on $\Pp\Pp$ data.  The value of \RAA as a function of $z$ shows a rising trend.  The suppression at low $z$ is found to be larger in the 20\% most central events (\ie ``head-on'' collisions), as compared to the less central selection. The results show explicitly that the \JPsi produced with a large degree of surrounding jet activity are more highly suppressed than those produced in association with fewer particles.  This finding emphasizes the importance of incorporating the jet quenching mechanism in models of \JPsi suppression.

\begin{acknowledgments}

  We congratulate our colleagues in the CERN accelerator departments for the excellent performance of the LHC and thank the technical and administrative staffs at CERN and at other CMS institutes for their contributions to the success of the CMS effort. In addition, we gratefully acknowledge the computing centers and personnel of the Worldwide LHC Computing Grid and other centers for delivering so effectively the computing infrastructure essential to our analyses. Finally, we acknowledge the enduring support for the construction and operation of the LHC, the CMS detector, and the supporting computing infrastructure provided by the following funding agencies: BMBWF and FWF (Austria); FNRS and FWO (Belgium); CNPq, CAPES, FAPERJ, FAPERGS, and FAPESP (Brazil); MES (Bulgaria); CERN; CAS, MoST, and NSFC (China); MINCIENCIAS (Colombia); MSES and CSF (Croatia); RIF (Cyprus); SENESCYT (Ecuador); MoER, ERC PUT and ERDF (Estonia); Academy of Finland, MEC, and HIP (Finland); CEA and CNRS/IN2P3 (France); BMBF, DFG, and HGF (Germany); GSRT (Greece); NKFIA (Hungary); DAE and DST (India); IPM (Iran); SFI (Ireland); INFN (Italy); MSIP and NRF (Republic of Korea); MES (Latvia); LAS (Lithuania); MOE and UM (Malaysia); BUAP, CINVESTAV, CONACYT, LNS, SEP, and UASLP-FAI (Mexico); MOS (Montenegro); MBIE (New Zealand); PAEC (Pakistan); MSHE and NSC (Poland); FCT (Portugal); JINR (Dubna); MON, RosAtom, RAS, RFBR, and NRC KI (Russia); MESTD (Serbia); SEIDI, CPAN, PCTI, and FEDER (Spain); MOSTR (Sri Lanka); Swiss Funding Agencies (Switzerland); MST (Taipei); ThEPCenter, IPST, STAR, and NSTDA (Thailand); TUBITAK and TAEK (Turkey); NASU (Ukraine); STFC (United Kingdom); DOE and NSF (USA).
   
  \hyphenation{Rachada-pisek} Individuals have received support from the Marie-Curie program and the European Research Council and Horizon 2020 Grant, contract Nos.\ 675440, 724704, 752730, 765710 and 824093 (European Union); the Leventis Foundation; the Alfred P.\ Sloan Foundation; the Alexander von Humboldt Foundation; the Belgian Federal Science Policy Office; the Fonds pour la Formation \`a la Recherche dans l'Industrie et dans l'Agriculture (FRIA-Belgium); the Agentschap voor Innovatie door Wetenschap en Technologie (IWT-Belgium); the F.R.S.-FNRS and FWO (Belgium) under the ``Excellence of Science -- EOS" -- be.h project n.\ 30820817; the Beijing Municipal Science \& Technology Commission, No. Z191100007219010; the Ministry of Education, Youth and Sports (MEYS) of the Czech Republic; the Deutsche Forschungsgemeinschaft (DFG), under Germany's Excellence Strategy -- EXC 2121 ``Quantum Universe" -- 390833306, and under project number 400140256 - GRK2497; the Lend\"ulet (``Momentum") Program and the J\'anos Bolyai Research Scholarship of the Hungarian Academy of Sciences, the New National Excellence Program \'UNKP, the NKFIA research grants 123842, 123959, 124845, 124850, 125105, 128713, 128786, and 129058 (Hungary); the Council of Science and Industrial Research, India; the Latvian Council of Science; the Ministry of Science and Higher Education and the National Science Center, contracts Opus 2014/15/B/ST2/03998 and 2015/19/B/ST2/02861 (Poland); the National Priorities Research Program by Qatar National Research Fund; the Ministry of Science and Higher Education, project no. 0723-2020-0041 (Russia); the Programa Estatal de Fomento de la Investigaci{\'o}n Cient{\'i}fica y T{\'e}cnica de Excelencia Mar\'{\i}a de Maeztu, grant MDM-2015-0509 and the Programa Severo Ochoa del Principado de Asturias; the Thalis and Aristeia programs cofinanced by EU-ESF and the Greek NSRF; the Rachadapisek Sompot Fund for Postdoctoral Fellowship, Chulalongkorn University and the Chulalongkorn Academic into Its 2nd Century Project Advancement Project (Thailand); the Kavli Foundation; the Nvidia Corporation; the SuperMicro Corporation; the Welch Foundation, contract C-1845; and the Weston Havens Foundation (USA).

\end{acknowledgments}

\bibliography{auto_generated}

\cleardoublepage \appendix\section{The CMS Collaboration \label{app:collab}}\begin{sloppypar}\hyphenpenalty=5000\widowpenalty=500\clubpenalty=5000\input{HIN-19-007-authorlist.tex}\end{sloppypar}
\end{document}

%% file: HIN-19-007-authorlist.tex
\vskip\cmsinstskip
\textbf{Yerevan Physics Institute, Yerevan, Armenia}\\*[0pt]
A.~Tumasyan
\vskip\cmsinstskip
\textbf{Institut f\"{u}r Hochenergiephysik, Wien, Austria}\\*[0pt]
W.~Adam, T.~Bergauer, M.~Dragicevic, J.~Er\"{o}, A.~Escalante~Del~Valle, R.~Fr\"{u}hwirth\cmsAuthorMark{1}, M.~Jeitler\cmsAuthorMark{1}, N.~Krammer, L.~Lechner, D.~Liko, T.~Madlener, I.~Mikulec, F.M.~Pitters, N.~Rad, J.~Schieck\cmsAuthorMark{1}, R.~Sch\"{o}fbeck, M.~Spanring, S.~Templ, W.~Waltenberger, C.-E.~Wulz\cmsAuthorMark{1}, M.~Zarucki
% ...
\vskip\cmsinstskip
% ...
1:  Also at TU Wien, Wien, Austria\\
% ...
